\documentclass[sigconf]{acmart}
\usepackage[ruled,linesnumbered,boxed]{algorithm2e}
\usepackage{multirow}
\usepackage[flushleft]{threeparttable}
\usepackage{bm}
\AtBeginDocument{%
  \providecommand\BibTeX{{%
    \normalfont B\kern-0.5em{\scshape i\kern-0.25em b}\kern-0.8em\TeX}}}

\copyrightyear{2019}
\acmYear{2019}
\setcopyright{acmlicensed}
\acmConference[SALMM '19]{1st International Workshop on Search as Learning with Multimedia Information}{October 21, 2019}{Nice, France}
\acmBooktitle{1st International Workshop on Search as Learning with Multimedia Information (SALMM '19), October 21, 2019, Nice, France}
\acmPrice{15.00}
\acmDOI{10.1145/3347451.3356731}
\acmISBN{978-1-4503-6919-0/19/10}

\begin{document}
\fancyhead{}
%

\title{Investigating Correlations of Automatically Extracted Multimodal Features and Lecture Video Quality}

\author{Jianwei Shi}
\email{jianwei.shi.cn@outlook.com}
\orcid{0000-0003-0226-3608}
\affiliation{%
    \institution{L3S Research Center, Leibniz Universit\"at Hannover}
}

\author{Christian Otto}
\email{christian.otto@tib.eu}
\orcid{0000-0003-0226-3608}
\affiliation{%
    \institution{Leibniz Information Centre for Science and Technology~(TIB)}
}

\author{Anett Hoppe}
\email{anett.hoppe@tib.eu}
\orcid{0000-0003-0918-6297}
\affiliation{%
    \institution{Leibniz Information Centre for Science and Technology~(TIB)}
}

 \author{Peter Holtz}
 \email{p.holtz@iwm-tuebingen.de}
 \orcid{0000-0001-7539-6992}
 \affiliation{%
     \institution{Leibniz Institut f\"ur Wissensmedien~(IWM)}
 }

\author{Ralph Ewerth}
\email{ralph.ewerth@tib.eu}
\orcid{0000-0003-0918-6297}
    \affiliation{%
    \institution{Leibniz Information Centre for Science and Technology~(TIB)}
    \institution{L3S Research Center, Leibniz Universit\"at Hannover}
}

\renewcommand{\shortauthors}{Jianwei Shi, Christian Otto, Anett Hoppe, Peter Holtz, Ralph Ewerth}
\renewcommand{\shortauthors}{Jianwei Shi et al.}

\begin{abstract}
Ranking and recommendation of multimedia content such as videos is usually realized with respect to the relevance to a user query. However, for lecture videos and MOOCs (Massive Open Online Courses) it is not only required to retrieve relevant videos, but particularly to find lecture videos of high quality that facilitate learning, for instance, independent of the video's or speaker's popularity. Thus, metadata about a lecture video's quality are crucial features for learning contexts, e.g., lecture video recommendation in search as learning scenarios. In this paper, we investigate whether automatically extracted features are correlated to quality aspects of a video. A set of scholarly videos from a Mass Open Online Course (MOOC) is analyzed regarding audio, linguistic, and visual features. Furthermore, a set of cross-modal features is proposed which are derived by combining transcripts, audio, video, and slide content. A user study is conducted to investigate the correlations between the automatically collected features and human ratings of quality aspects of a lecture video. Finally, the impact of our features on the knowledge gain of the participants is discussed.
\end{abstract}

\keywords{multimodal, video assessment, correlation, knowledge gain}

\maketitle

\section{Introduction}
\label{sec:introduction}
Modern studying has left the classrooms -- more and more, learners refer to the Web as a source of educational material for a certain learning need. This kind of informal learning scenario is, nowadays, only insufficiently supported by Web mechanisms such as search engines which focus on satisfying information needs instead (\cite{baeza2011}). Recently, the research area \textit{Search as Learning} (SaL) gains momentum: It recognizes learning as an implicit component of Web search \cite{agosti2013}, aims to detect learning needs and intents \cite{eickhoff2014, vakkari2016} and to identify factors correlated to successful learning outcomes (e.~g., \cite{collins2016}).
While there are first interesting insights, the research field is still mostly geared to the recommendation of textual learning resources \cite{hoppe2018}. This stands in contrast to current learning research which suggests that users may prefer images and videos when tackling certain learning needs (e.~g., procedural learning tasks \cite{genuchten2012,pardi2019}). In consequence, this work sets out to establish a relationship between automatically extracted video features and successful learning. The objective is to facilitate more effective search for learning objects, be it for learners with a certain learning intent or for educators which seek for high-quality material to enhance their own courses. 
Normally, there is a large amount of educational and lecture videos that cover similar content -- from single-video tutorials to full-fledged Massive Open Online Courses (MOOCs). In such a large lecture video archive, it is desirable to find content of high quality regarding the knowledge presentation. However, objective features which assess, for instance, the nature of the presenters voice or the design of the slides, are not fully explored in automated systems in the current state of research. 
A human viewer evaluates the quality of a learning resource based on all available information. In common lecture videos this includes the textual, oral and visual modality. Viewers are supported in their learning by the visual elements on the slide, the words spoken and the gestures of the lecturer. 

Previous work, like Guo et al.~\cite{guo2014how}, investigated how the design of lecture videos impacts the viewer engagement and provided recommendations to optimize the content accordingly. Chen et al.~\cite{Chen2014} used multimodal sensing to assess the quality of a presentation. They extracted speech, body movement and visual features from the shown slides. Principal Component Analysis was applied to human ratings in order to address the two main modalities of the presentation: 1) recital skills, including, for instance, voice information and body language, and 2) slide quality, with regards to grammar, readability, and visual design. Pearson correlation was used to measure the relation between the different features.
Haider et al.~\cite{Haider2016} proposed a system for automatic video quality assessment, which is the most similar to our approach, focusing on prosodic and visual features. They extracted the complete set of audio features from the ComParE challenge~\cite{Schuller2013} and a total of 42 features related to hand movements of the speaker. The employed Multimodal Learning Analytics (MLA) dataset~\cite{Ochoa2014} contains 416 oral presentations (in Spanish) and the respective metadata regarding speech, facial expressions, skeletal data extracted from a Microsoft Kinect, as well as the shown slides. Each of these videos was labeled with ten individual ratings and an overall score related to the quality of the slides. A correlation study (discriminant analysis) was employed which found that prosodic features are able to predict \textit{self confidence and enthusiasm} (of the speaker) as well as \textit{body language and pose}, which is a quality measure their participants had to label. Their visual features showed similar results, but with less accuracy. 

In this paper, we go beyond previous work by (1) proposing a novel set of intuitive unimodal and cross-modal features that do not rely on skeletal data which are hard to acquire, (2) conducting an empirical evaluation on the correlation of features with quality aspects of a video, and (3) conducting an empirical evaluation on the correlation of features with participants' knowledge gain.  
Videos of a MOOC website are utilized and automatically annotated with our set of unimodal and multimodal features that address different quality aspects. Different modalities are exploited: audio and spoken text, video frame content, slide content, as well as cross-modal features. 
The experimental results reveal that several of our features show a significant correlation with the corresponding human assessment of the lecture videos. 

The remainder of the paper is organized as follows. Section 2 describes the set of extracted unimodal as well as novel cross-modal features. The design of the user study and the experimental results are presented in Section 3, while Section 4 concludes the paper and outlines areas of future work.
\section{Extraction of Unimodal and Cross-modal Features}
\label{sec:methodology}

\begin{figure}
	\centering
	\includegraphics[width=0.4\textwidth]{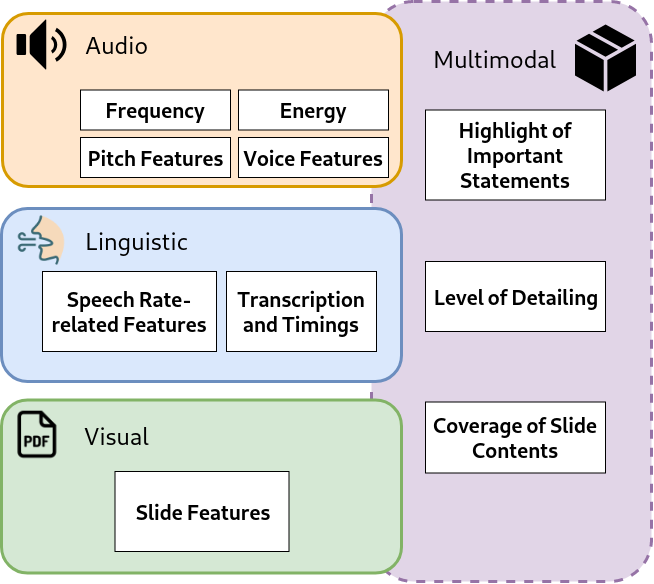}
	\caption{Overview over the feature sets extracted for the automatic assessment algorithm.}	
	\label{fig:overview-features}	
\end{figure}

This section outlines our approach for the extraction of in total 22 features from lecture videos including textual, audio, linguistic, as well as a set of chosen multimodal features.
An overview of our feature set is depicted in Figure~\ref{fig:overview-features}.
Since we are dealing with educational videos, we assume that for each data sample a video file is available, together with a PDF file of the shown presentation as well as a speech transcript. 

\subsection{Unimodal Features}
There are three kinds of unimodal features: audio, linguistic, and visual.
\subsubsection*{Audio Features}
The \textsf{openSMILE}-toolkit~\cite{Eyben2013} is used to extract the audio features, except for the pitch variation information which is extracted according to Hincks~\cite{Hincks2005}. 
We selected the feature subset of the ComParE challenge ($6.373$ dimensions, around 70 low-level descriptors with multiple features each) and reduced it to nine features and their arithmetic mean, see Table \ref{tab:audio_features}. For our study, we selected those features that have either shown an impact on the audio quality before (Jitter, F$0$ Harmonics ratio, ...) or are very likely to influence the perceived quality of the audio (Energy, Loudness, Harmonics-to-Noise Ratio, etc.). 

\begin{table}[h]
	\begin{tabular}{|p{2.5cm}|p{5.5cm}|}
	    \hline
	    \textbf{Feature} & \textbf{Description}\\
		\hline
	    Loudness                &  Sum of a simplified auditory spectrum   \\
		\hline
	    Modulated loudness      &  Sum of a simplified RASTA-filtered auditory spectrum~(RelAtive Spectral TrAnsform, Hermansky et al.~\cite{hermansky1992})   \\
		\hline
	    Root-Mean-Square energy &  Square root of mean of the discrete values of the sound pressure  \\
		\hline
	    Jitter                  &  Deviation from true periodicity of a presumably periodic signal\\
		\hline
	    $\Delta$~Jitter         &  Normalized average length deviation from true periodicity of a presumably periodic signal  \\
		\hline
	    Shimmer                 &  Amplitude variation of consecutive voice signal periods\\
		\hline
	    Harmonicity (spectral)  &  Ratio between the minima and the maxima in relation to the amplitude of the maxima from a magnitude spectrum\\
		\hline
	    Logarithmic Harmonics-to-Noise Ratio    & Logarithmic scale of the ratio harmonic to noise component in the wave signal\\
		\hline
	    Pitch Variant Quotient    & Standard deviation of the pitch, which is divided by the mean of the pitch (cf. Hincks~\cite{Hincks2005})\\
		\hline
	\end{tabular}
	\caption {Automatically extracted audio features.} \label{tab:audio_features} 
\end{table}

\subsubsection*{Linguistic Features}
The extraction of linguistic features aims to describe how the presenters articulate themselves by means of syllable duration and speaking rate. De Jong and Wempe's~\cite{DeJongWempe2009} \textsf{Praat}~\cite{web:praat} script was used to extract these features, namely \textit{speech rate}, \textit{articulation rate}, and average syllable duration~(\textit{ASD}). All of them are derived from the number of vowels or syllables per time interval and indicate if the speaker is talking too fast or too slow.

The video transcript does also contain a lot of useful information regarding speech quality. Since the influence of textual content on the knowledge gain has been examined extensively (e.g., by ~\citet{Gadiraju2018analyzing}), this information is disregarded here and the focus set on the lesser researched features.
However, we make use of the content of the transcript in a different way in section~\ref{sec:cross_modal}.
\subsubsection*{Visual Features}
\label{sec:visual_features}
For the visual content, we examine the PDF files of the presentation slides. With the bash command \textsf{pdftotext} we extract text layout information from the PDF slides. The command extracts the position of each text element as well as the size of the slide. The elements of a slide are stored in an hierarchical way starting with the biggest text element, which contains multiple text lines and each line consists of multiple words. This information is converted into a XHTML file. Similarly, the \textsf{pdftohtml} command is used to extract the image position and size of the slide, which are stored in an XML file. The generated files are then parsed to JSON, since the format is more convenient for data handling.

Based on this representation, we compute two features related to the design of the slides, which are \textit{text ratio} and \textit{image ratio}. They describe, how much slide space is covered by each of the modalities according to Formula~\ref{eq:text_ratio}. 
\begin{equation}
\label{eq:text_ratio}
Text Ratio=\dfrac{\sum_{i=1}^n TextArea_i}{Area_{slide}}.
\end{equation}
Also, for each file we store the mean and sample variance of the text ratio and image ratio values of all slides.

\subsection{Cross-modal Features}
\label{sec:cross_modal}
In this section, we present a set of multimodal features which aim to model specific quality aspects of a presentation. We were inspired by criteria that are important to us humans, for instance, the way and frequency the presenter highlights important aspects on the slides. If we are able to capture these metrics, we can rank videos with similar content according to their presentation quality providing an optimal recommendation to a learner. 

\subsubsection*{\textbf{Highlight of Important Statements}}

This feature is supposed to indicate how often important statements are emphasized per slide and over the complete slide set. To identify the text boxes most likely containing the key components of a slide, we use the information from the document layout analysis, which we stored in JSON format earlier.

In this procedure, we use the following natural language processing functions which were adapted from Bird et al.~\cite{Bird2009}:
\begin{itemize}
	\item N():   return a list of nouns from a sentence
	\item LEM(): return a set of lemmas from a list of words
	\item SYN(): return a set of synonyms from a list of words
\end{itemize} 

Our assumption for the identification of \textit{important} text is that font size is often proportional to importance. Since we do not have the font size information for each slide we sort the text lines by the area they cover. However, simply choosing the \textit{n} largest text areas does not yield good results, because there are often bullet points of similar importance  but cover a text areas of different size. So we cluster the text areas according to the following rule: For each text area starting from the biggest one, if the area difference to the next biggest text area is smaller than $n\%$ of the slide size (in our experiments $1\%$), add it to the existing cluster, otherwise create a new one. All text areas in the two biggest clusters are considered to contain important statements. This usually includes the text area of the title and all headlines of the highest category.
For each selected text line, we first extract the sentence(s) ($St_{imp}$) from the respective JSON file. Then, the nouns and their synonyms are extracted from the text. Finally, we lemmatize the nouns and their synonyms:
\begin{equation}
St = LEM(N(St_{imp}) \cup SYN(N(St_{imp}))).
\end{equation}
\subsubsection*{\textbf{Locating Emphasized Transcriptions}}
\begin{figure}[!ht]
	\centering
	\includegraphics[width=0.5\textwidth]{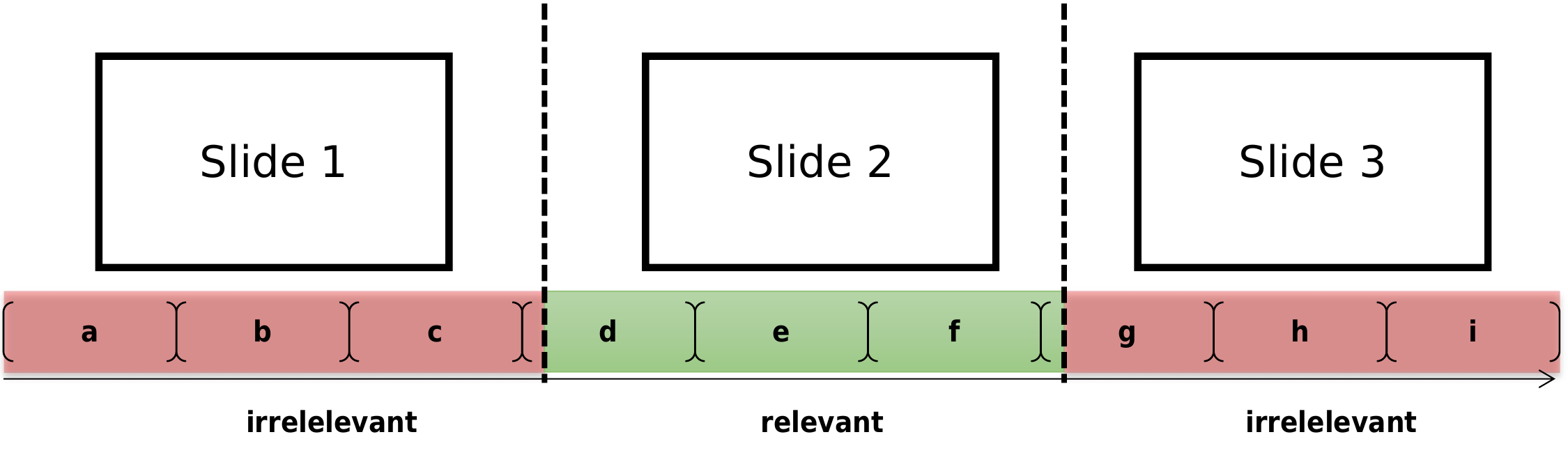}
	\caption{Visualization of the overlap between the speech transcript blocks $a, ..., i$ and the slides $1-3$. Based on their percental overlap with the slide, a certain percentage of words is removed from each block.}	
	\label{fig:slide_problem}	
\end{figure}

This feature is designed to capture the ability of the presenter to consider important statements shown on the slides as well as their emphasis through his or her voice. If so, there should be a corresponding local maximum in the audio signal. To get this information, we need to align the speech transcript with the audio signal in the time frame where the currently observed slide was covered. The segmentation of the video according to  single slides is done manually. 
For each slide and associated time frame we need to find the corresponding segment of the speech transcript. The transcript is segmented into blocks of 10 seconds, which is a common interval in speech analysis (cf. Hincks~\cite{Hincks2005}). A slide segment of arbitrary length can contain multiple of these blocks and it is important to find the correct one for each highlighted statement, see Figure~\ref{fig:slide_problem}. This is done using the audio signal.
We encoded the audio information via three metrics: F$_{0} $, loudness, and energy. If all of them have a maximum at around the same timestamp, we assume the presenter emphasized the word said at that moment. Locating the exact words at that moment is done by choosing the speech transcript block whose temporal centre is closest to the found maximum. The list of these presumed important statements are stored in \textit{EmphasizedTranscriptions} and we lemmatize them again:
\begin{equation}
Tr = LEM(\textit{EmphasizedTranscriptions}).
\end{equation}

Finally, the fraction of highlighted statements is calculated as:
\begin{equation}
\label{eq:highlight}
Highlight = \dfrac{|St\cap Tr|}{|St|}.
\end{equation}

\subsubsection*{\textbf{Level of Detailing}}
Another possible measure of quality for the presentation is the overlap of spoken text with the information on the slide. As literature from video learning suggests \cite{brame2016}, audio and visual contents should provide complementary information rather than being overly redundant. Thus, we examine if the speaker only read the information already present on the slide, or if the oral explanation provides further detail, giving an appropriate amount of additional information. For this purpose, we calculate the ratio of said words to shown words on the slide.

Again, we use the speech transcript blocks from the previous metric to count the number of words said during the time frame when the corresponding slide was visible. Every speech transcript block overlapping with the duration of the slide is considered. Blocks at the interval boundaries are cut off appropriately. If a transcript block overlapped $70\%$ at the end of a slide and contained 10 words, we would consider the first seven words and would dismiss the last three. All found words are therefore stored in $Words_{said}$. The number of words on the slide $Words_{slide}$ were gathered from the process explained in section~\ref{sec:visual_features}.

The \textbf{level of detailing} is calculated as the ratio of number of said words to number of words on the slide, see Formula~\ref{eq:detail}:
\begin{equation}
\label{eq:detail}
LevelOfDetailing=\dfrac{|Words_{said}|}{|Words_{slide}|}.
\end{equation}

Also, for each video the mean and sample variance of these values was calculated for later usage.
\subsubsection*{\textbf{Coverage of Slide Contents}}
This metric encapsulates if the speaker talked about all the information shown on the slide of if some parts were skipped. Again, this gives an idea if the overall talk is well structured and timed or rushed and hard to follow for an observer, since some information shown is left without explanation. 
We can reuse the $Words_{slide}$ from the previous section. 
For the words said by the speaker, we reuse the already established $Words_{said}$ from the previous metric.

The words in $Words_{slide}$ and $Words_{said}$ were lemmatized again, which enables an easier comparison. The \textbf{Coverage} of slide content is calculated as the ratio of the number of common words in $Words_{said}$ and $Words_{slide}$ to the total number of words on the slide:
\begin{equation}
Coverage=\dfrac{|Words_{said} \cap Words_{slide}|}{|Words_{slide}|}.
\end{equation}
Similarly, the mean and sample variance of the values of all slides are calculated for later usage.
The full list of manual and automatic features can be seen in Table~\ref{tab:features}.

\begin{table}[]
\begin{tabular}{l|l}
\textbf{Human-rated Quality Aspect}  & \textbf{Automatic Features}    \\
\hline
Clear Language            & Loudness avg.                  \\
Vocal Diversity           & mod. Loudness avg.             \\
Filler Words              & RMS Energy avg.                \\
Speed of Presentation     & f0 avg.                        \\
Coverage of the Content   & Jitter avg.                    \\
Level of Detail           & $\Delta$ Jitter avg.           \\
Highlight of imp. Content & Shimmer avg.                   \\
Summary                   & Harmonicity avg.               \\
Text Design               & log. HNR avg.                  \\
Image Design              & PVQ avg.                       \\
Formula Design            & Speech Rate                    \\
Table Design              & Articulation Rate              \\
Structure of Presentation & avg. Syllable Duration         \\
Entry Level               & Text Ratio avg.                \\
Overall Rating            & Text Ratio var.                \\
                          & Image Ratio avg.               \\
                          & Image Ratio var.               \\
                          & Highlight of imp. Statements   \\
                          & Level of Detailing avg.        \\
                          & Level of Detailing var.        \\
                          & Coverage of Slide Content avg. \\
                          & Coverage of Slide Content var. \\
\end{tabular}
\vspace{0.1cm}
\caption{Overview of all recorded features.}
\label{tab:features}
\end{table}

\section{Design of the User Study}
\label{sec:experiments}
To estimate the expressiveness of our automatically extracted features we conducted a user study. We aimed to get human ratings for different quality aspects of lecture videos, while these aspects are covered by respective feature sets. In addition, we asked for an overall rating of a lecture video. Furthermore, every participant was asked to fill in a knowledge test before (pre-test) and after (post-test) watching the video, aiming to measure the capability of a video to convey knowledge. We conducted a correlation analysis to find out which features are correlated with the quality aspects of lecture videos as well as with knowledge gain.


\subsection{Data Acquisition}
\label{sec:data}
Our dataset consists of 22 videos (with associated slides and speech transcripts) from edX\footnote{\url{https://courses.edx.org/courses/course-v1:DelftX+GSE101x+1T2018/course/}}. The course materials of this course are Copyright Delft University of Technology and are licensed under a Creative Commons Attribution-NonCommercial-ShareAlike 4.0 International License~\cite{web:cc:4}. The available course materials on edX are provided the following formats: videos in MP4, slides in PDF, and transcriptions in SRT. We chose this source since it does not require any further pre-processing, it is open access and the speech transcript is of high quality (presumably they have been manually reviewed). 

\subsection{Participants and Task}
The subject of the 22 videos is software engineering. Each video has exactly one presenter, while the full dataset has nine different presenters with varying slide designs.
We employed 13 participants (10 men, 3 women) from our university with a computer science background, an average age of $25.8 \pm 2.4$ years, and asked everyone to watch and assess nine videos. With an average video length of 8 minutes and the time to fill out the evaluation forms the full experiment took $1.5 - 2$ hours. The participants were rewarded $13$ Euro/Hour. We made sure that each set of videos contained as many different presenters as possible and that each video is viewed at least by five different people. We chose to gather multiple ratings for the same video instead of investigating a larger set of videos to be more robust against outlier ratings.

\subsection{Experimental Setting}
A common way to estimate the knowledge gain of a participant during a learning session is to conduct a knowledge test before and after a controlled learning session (e.g., \citet{yu2018predicting}). The resulting score difference indicates how much was learned. Even though the potential knowledge gain depends heavily on the participant, by choosing a subject that is most likely unfamiliar to a majority of people, we try to circumvent this problem. It is however important to ensure that the participants have a chance to understand the content, otherwise the knowledge gain will be low again. Therefore, we selected the topic \textit{Globally Distributed Software Engineering}\footnote{https://www.edx.org/course/globally-distributed-software-engineering-2}. It is on one hand, a computer science topic related to the studies of our participants but also a very specific area which is not part of their curriculum. Thus, everyone had a chance to understand the topic based on their prior knowledge and therefore favouring a positive knowledge gain.

\begin{figure}
	\centering
	\fbox{
	    \includegraphics[width=0.45\textwidth]{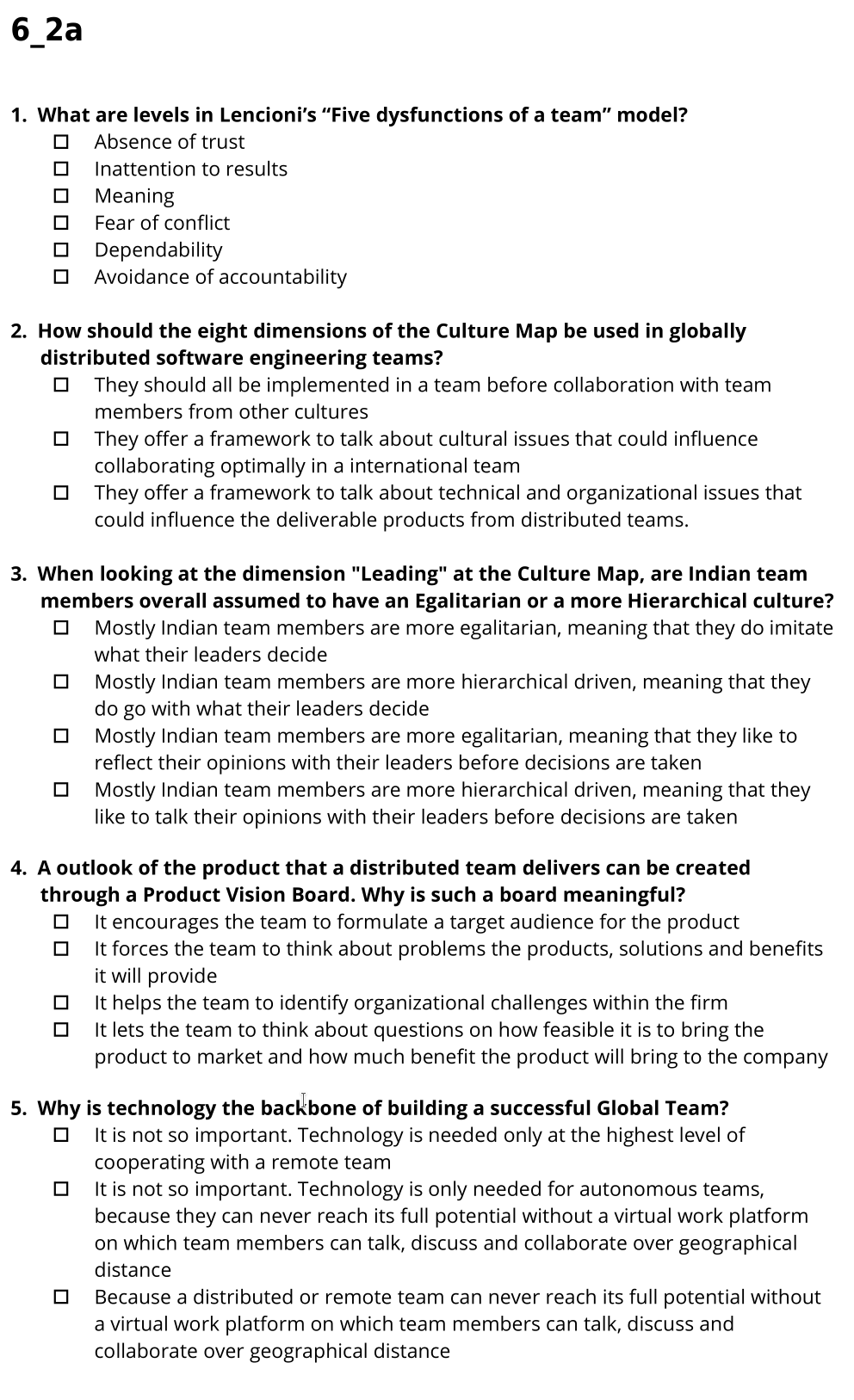}
	}
	\caption{The questionnaire for the pre- and post test of video 6\_2a. Questions 1 and 3 are relevant to this video.}
	\label{fig:kg_test}
\end{figure}

A negative effect of the pre-test is that it might influence the user behavior by providing hints on what to focus on in a video, because participants will try to get a good score on the post-test. We gathered a set of relevant questions inspired by the intermediate quizzes in the course material. However, we made sure to amend and change them since their reuse is restricted. We chose two to four questions intended to be asked after each video with a similar amount of unrelated questions from other videos. Also, we put the videos in random order so it was hard to guess which of the questions will be the relevant ones. In addition, the number of possible answers was different every time. Exemplary, the knowledge test for video 6\_2a can be seen in Figure~\ref{fig:kg_test}.

After filling out the pre-test, the participant was instructed to watch the entire video without pausing, rewinding, or taking notes. Reason for that is that we wanted the participants to get a full impression of the presentation instead of, again, just skipping to the relevant parts for the knowledge test to get a good score. Similarly, we assumed that when we allowed people to take notes, they would just write reminders down about the pre-test and focus solely on their appearance in the video. Admittedly, this is slightly different to a realistic setting, but we applied it in favor of the knowledge gain measurement. 
After watching the video, the person is asked to answer the same questions again and also to fill out an evaluation form with questions that are related to different quality aspects, see Table~\ref{tab:study_features}. The items are assessed using a Likert scale from 1-5. 
\begin{table}[ht]
	\centering
	\begin{tabular}{|p{3cm}|l|}
		\hline
		\textbf{Automatic features} &  \textbf{Human-rated aspects}  \\
		\hline
		\multirow{2}{1.5cm}{Audio}              &  Clear language                       \\
		                                        &  Vocal diversity                      \\
		\hline
		\multirow{2}{1.5cm}{Linguistic}         &  Filler words    \\
		                                        &  Speed of presentation    \\
		\hline
		\multirow{2}{1.5cm}{Visual}             &  text/image/formula/table design \\
	                                            &  Structure of the presentation    \\
		\hline
		                                        & Coverage of the slide content  \\
                                                & Appropriate level of detail   \\
         Cross-modal                            & Highlight of important content     \\
                                                & Summary        \\
                                                & Overall rating   \\
		\hline
	\end{tabular}
    \caption {Automatically extracted features and corresponding items in the evaluation form of the user study.} \label{tab:study_features} 
\end{table}

\begin{figure}
	\centering
	\includegraphics[width=0.45\textwidth]{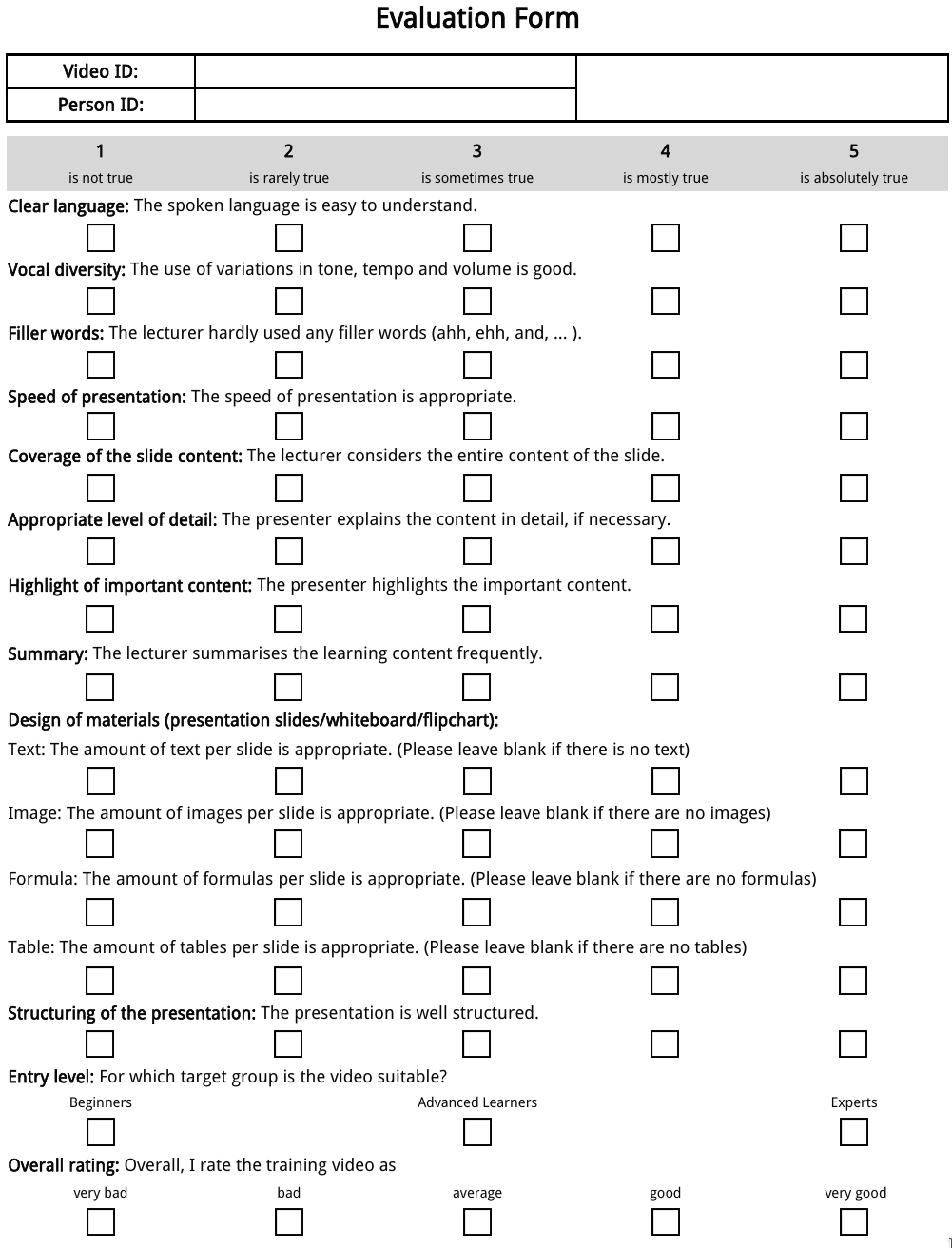}
	\caption{The full evaluation form the users had to fill out for each video.}
	\label{fig:evaluation-form}
\end{figure}

This paragraph describes how we scored the knowledge test with the goal of treating each video with a similar importance, independent of 1) the number of relevant questions per quiz and 2) the number of possible answers per question.
First, the score for an unanswered question will be treated as zero since we gave the participants the option to skip a question in order to discourage random guessing.
If the question was answered, we calculate the score for each answer option by increasing (decreasing) the score by 1 for a correct (false) answer. Thus, a question with five answer options can yield the following scores: ${-5, -3, -1, 0, 1, 3, 5}$. 


The knowledge gain of participant $s$ after watching video $v$ is then calculated as the difference between the pre-test score $PB_{vs}$ and post-test score $PA_{vs}$. Let $n_{v}$ be the number of participants who watched video $v$.

We start by computing the mean knowledge gain of participants for each video:\begin{equation}
\mu = \dfrac{\sum_{j=1}^{n_{v}}(PA_{vs} - PB_{vs}) }{2n_{v}}.
\end{equation}
Next is the standard deviation of the knowledge gain:
\begin{equation}
\sigma = \sqrt{\frac{1}{2n_{v}}[\sum_{s=1}^{n_{v}}(PB_{vs}-\mu)^{2}+\sum_{s=1}^{n_{v}}(PA_{vs}-\mu)^{2}]}.
\end{equation} 
Based on the mean and standard deviation value, the scores are normalized.
$PB_{vs}^{'}$ is the normalized score which is computed by subtracting the mean and dividing by the standard deviation. The same applies for $PA_{vs}^{'}$:
\begin{equation}
PB_{vs}^{'} = \dfrac{PB_{vs}-\mu}{\sigma}.
\end{equation}
Consequently, the normalized knowledge gain of participant $s$ for video $v$ is:
\begin{equation}
KG_{vs} = PA_{vs}^{'} - PB_{vs}^{'}.
\end{equation}
$KG_{v}$, the overall knowledge gain for video $v$ is finally calculated by the average of all participants' knowledge gain  :
\begin{equation}
KG_{v} = \dfrac{1}{n_{v}}\sum_{s=1}^{n_{v}} KG_{vs}.
\end{equation}

\section{Results and Discussion}
\label{sec:results}
We conducted a correlation analysis to compare the automatically extracted features with their human-labeled counterparts. The following section discusses the results of this analysis.

\subsection{Correlation Measures}
For each pair of features we compute a correlation coefficient $r$ and a confidence value $\alpha$ (assuming a two-tailed hypothesis). $r$ is in the range of $[-1, 1]$. $r=-1$ indicates a strong negative correlation, $r=0$ for uncorrelated data and $r=+1$ indicates a strong positive correlation. 
The threshold $\alpha$ represents how extreme the observed correlation has to be in order to reject the null hypothesis (''The correlation happened by chance.``). 
The $\alpha$ values are set according to the table of exact critical values for Spearman's correlation coefficient by Ramsey~\cite{Ramsey1989}. For Pearson's correlation coefficient, the $\alpha$ thresholds are given by Weathington et al.~\cite{Weathington2012} (p.~452) and an extension table which is calculated according to the description in ~\cite{Weathington2012} (p.~451) using an R script~\cite{RProject}.

In order to apply the appropriate correlation coefficient $r$ together with the correct $\alpha$ value, multiple aspects have to be considered. First, based on the data type of the analyzed feature pair we have to decide between Spearman's or Pearson's correlation coefficient. Spearman's coefficient is applicable to the majority of combinations, because all the automatic features are interval data and the manually labeled features are entirely ordinal data, except for the results of the knowledge gain test. The difference between post- and pre-knowledge tests are interval scaled, therefore Pearson analysis is the appropriate method to conduct the correlation analysis between the knowledge gain score and the automatic features.

The next important parameter to determine is the degree of freedom $df = N - 2$, where $N$ is the sample size. 
There are 22 videos for correlation analysis, so for the following analysis $N = 22$ and $df = 20$.

\subsection{Correlations of Features and Quality}
For each video, we have at least five annotations from which we compute the mean value. Table~\ref{tab:pos-spearman-cor} shows the correlation results with a confidence level $\alpha < 0.05$, so the effect is not coincidental with a confidence of at least $95\%$. Similarly, Table~\ref{tab:neg-spearman-cor} shows negative correlation results. The following subsections discuss the subset of human labeled quality measures that had a statistically significant correlation with at least two automatically extracted features.
\begin{table}[h]
	\centering
	\newlength{\autowidth}
	\setlength{\autowidth}{3.5cm}
	\begin{threeparttable}
	\begin{tabular}{|c|c|c|c|}
		\hline
		\textbf{Human-rated Aspect} & \textbf{Autom. Features}   &\bm{$ r_{s} $} & \boldmath{$ \alpha < $} \\ \hline \hline
		\multirow{6}{*}{Clear Language}       & RMS Energy                    & 0.368                                     & 0.1  \\ 
		                                      & Loudness                      & 0.430                                     & 0.05 \\  
		                                      & Harmonicity (spectral)        & 0.435                                     & 0.05 \\ 
	                                          & Detailing Mean                & 0.541                                     & 0.02 \\ 
	                                          & Detailing Variance            & 0.448                                     & 0.05 \\
		                                      & $F_0$                         & 0.608                                     & 0.005 \\ 
		\hline
		\multirow{6}{*}{Vocal Diversity}      & Log. HNR                      & 0.389                                     & 0.01 \\  
		                                      & $F_0$                         & 0.429                                     & 0.05 \\
	                                          & Detailing Variance            & 0.463                                     & 0.05 \\ 
		                                      & Speech Rate                   & 0.525                                     & 0.02 \\
	                                          & Detailing Mean                & 0.615                                     & 0.005 \\
		\hline
    	Summary                               & $F_0$                         & 0.420                                     & 0.1  \\ 
		\hline
		\multirow{2}{*}{Speed of Presentation}& $\Delta$ Jitter               & 0.408                                     & 0.1  \\
		                                      & Jitter                        & 0.423                                     & 0.1  \\ 
		\hline
		\multirow{4}{*}{Filler Words}         & Image Ratio Mean              & 0.394                                     & 0.1  \\
		                                      & Articulation Rate             & 0.459                                     & 0.05 \\ 
	                                          & Detailing Variance            & 0.493                                     & 0.05 \\ 
	                                          & Detailing Mean                & 0.601                                     & 0.005 \\ 
		\hline
		Image Design                          & ASD                           & 0.371                                     & 0.1  \\ 
		\hline
		Appropriate Detailing                 & Detailing Variance            & 0.378                                     & 0.1  \\  
		\hline
	\end{tabular}
    \caption {Positive correlation results with $\alpha < 0.1$ ($ r_{s} $: Spearman Correlation Coefficient, HNR: Harmonics-to-Noise Ratio, Detailing: Level of Detailing)} 
    \label{tab:pos-spearman-cor} 
    \end{threeparttable}
\end{table}
\begin{table}[h]
	\centering
	\begin{threeparttable}
		\begin{tabular}{|c|c|c|c|}
			\hline
			\textbf{Human-rated Aspect} & \textbf{Automatic Features}   & \bm{$ r_{s} $} & \boldmath{$ \alpha < $} \\ \hline \hline
			\multirow{3}{*}{Clear Language}                  & PVQ Average                 & -0.416 & 0.1   \\ 
			                                                 & Coverage Variance           & -0.489 & 0.05  \\ 
			                                                 & Shimmer                     & -0.623 & 0.005 \\
			\hline
			\multirow{3}{*}{Vocal Diversity}                 & Text Ratio Mean             & -0.396 & 0.1   \\  
			                                                 & Highlight                   & -0.406 & 0.1   \\
			                                                 & Shimmer                     & -0.437 & 0.05  \\
			\hline
			Overall Rating                                   & Speech Rate                 & -0.455 & 0.05  \\
			\hline
			Image Design                                     & Articulation Rate           & -0.368 & 0.1   \\ 
			\hline
			\multirow{2}{*}{Speed of Presentation}           & Text Ratio Variance         & -0.454 & 0.05  \\ 
			                                                 & Image Ratio Variance        & -0.561 & 0.01  \\ 
			\hline
			Text Design                                      & Image Ratio Mean            & -0.466 & 0.05  \\ 
			\hline
			Structure                                        & Coverage Variance           & -0.402 & 0.1   \\ 
			\hline
			Summary                                          & Coverage Variance           & -0.457 & 0.05  \\ 
            \hline
		\end{tabular}
	\caption {Negative correlation results with $\alpha < 0.1$ ($ r_{s} $: Spearman Correlation Coefficient, Coverage: Coverage of Slide Contents, Structure: Structuring of Presentation)}
	\label{tab:neg-spearman-cor} 
	\end{threeparttable}
\end{table}

\subsubsection*{\textbf{Clear Language}}
Several of our automatically extracted features from the audio signal have a strong correlation ($\alpha<0.025$) with the manual rating for \textit{clear language}, especially \textit{loudness}, \textit{RMS energy}, \textit{harmonicity (spectral)}, and $F_{0}$. This is intuitive according to their definitions in Table~\ref{tab:audio_features} and shows that this quality feature of a video can most likely be determined automatically when these features are regarded jointly.
With respect to negative correlations, there is a strong connection ($\alpha <0.0025$) between \textit{shimmer} and  \textit{clear language}. It correlates with \textit{vocal diversity} ($\alpha <0.025$) as well. This finding confirms that shimmer is an indicator about sickness in the voice (cf. Teixeira et al.~\cite{Teixeira2013}), which impairs the quality of the language. Unfortunately, the negative correlation with \textit{PVQ average} contrasts with Hincks~\cite{Hincks2005} findings that a higher PVQ score is associated with a more lively voice. This inconsistency may be due to the data smoothing for F$_{0}$ values. The smoothed values are less accurate and could lead to this effect.

\subsubsection*{\textbf{Vocal Diversity}}
The \textit{Vocal Diversity} of the speaker had a strong correlation with three audio features and two multimodal features. While the latter ones lack an intuitive explanation, the features from the audio signal, namely \textit{logarithmic HNR}, $F_{0}$, and \textit{speech rate}, most certainly directly contributed to this measurement. For instance, Yumoto et al.~\cite{Yumoto1982} found a negative correlation between Harmonics-to-Noise Ratio and hoarseness, hinting that this metric resembles the voice quality of the speaker. For the negative correlations, \textit{shimmer}, as mentioned before, severely impacts this metric and should definitely be considered when trying to automatically predict it.

\subsubsection*{\textbf{Speed of Presentation}}
The rating regarding an appropriate presentation speed positively correlated with the two \textit{jitter} features, which indicates changes in the speech rate. Possibly, the presenter regularly adapted the presentation speed based on the difficulty of the current slide -- which would be an indicator for a good presentation.
It is negatively correlated to the amount of images on the slides, which can be explained by the fact that an excessive amount of images on slides adds to the cognitive load: without sufficient time to process visual contents, the viewer might feel overwhelmed.

\subsubsection*{\textbf{Filler Words}}
Our \textit{filler words} quality metric has a strong correlation with the \textit{articulation rate} feature. This can be explained by the fact that a high articulation rate represents a high amount of syllables per time unit, which is especially apparent if the speaker uses a lot of filler words like ``uhhhmm'' or ``eehmm''. The remaining features associated with it do not allow for a direct interpretation but might be correlated indirectly. 

\subsubsection*{\textbf{Appropriate Detailing}}
Our cross-modal feature \textit{appropriate detailing} has a strong correlation with the variance of its automatic counterpart, namely \textit{detailing variance} ($\alpha < 0.05$). This is interesting since it indicates that the participants valued a high variance in the detailed parts of the speech more than a presentation where it happened constantly. In other words, at least for our dataset, a good presenter should only detail the necessary sections of the content.

\subsection{Correlation analysis for features and knowledge gain}
One of the most important features for an educational video is its ability to convey the intended information and, consequently, to improve the viewers' knowledge state. The following section highlights the automatically extracted features that had the biggest impact on the knowledge gain of the participants during the user study.
Since both our extracted features and the knowledge gain scores are interval data, Pearson correlation analysis is used.

\begin{figure}
	\centering
	\includegraphics[width=0.49\textwidth]{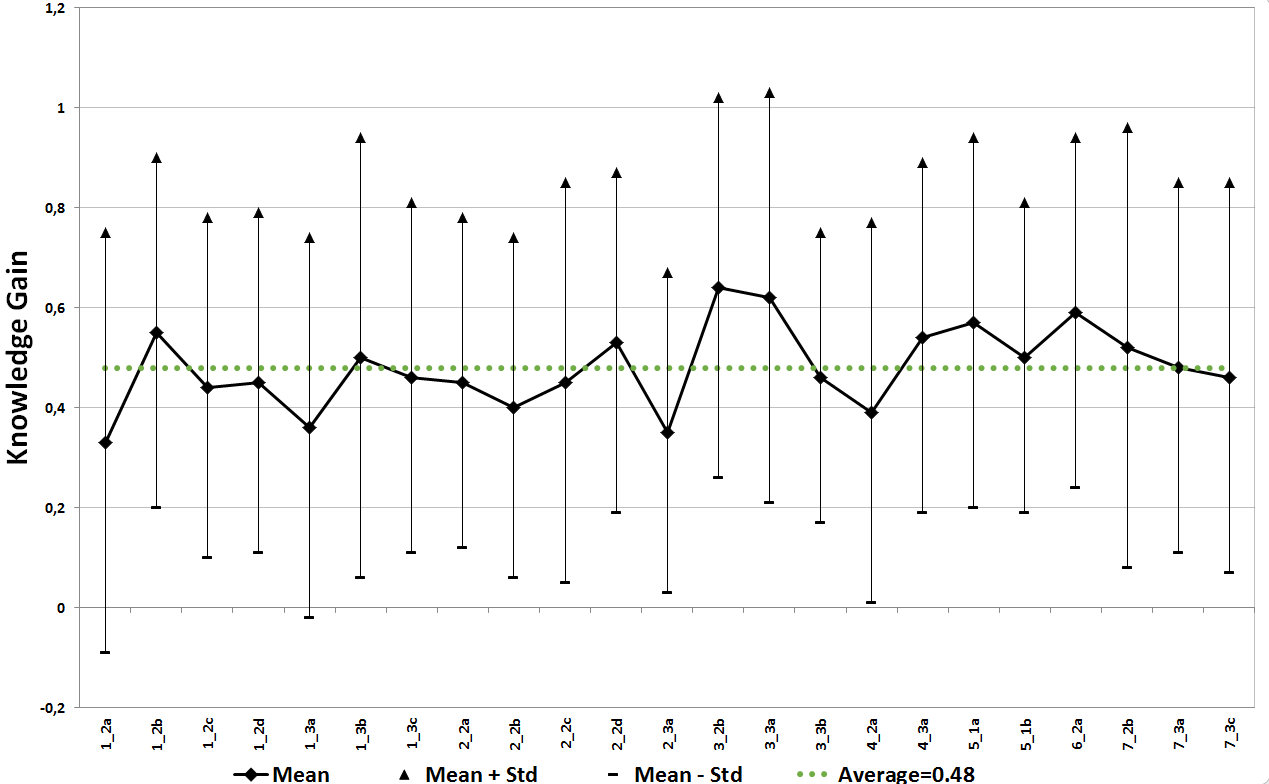}
	\caption{Knowledge gain distribution for all relevant questions in the user study (5 or 6 participants per question).}
	\label{fig:questions}
\end{figure}

First, Figure~\ref{fig:questions} shows the distribution of achieved knowledge gains for all 23 relevant questions answered during the user study. The data is normalised preventing questions with more possible answers to skew the results. The average knowledge gain of $+0.48$ shows that the chosen topic enabled all 
participants to improve their knowledge in this area to a certain degree. 
Also, the questions were neither to easy nor to hard, which would have also resulted
in a low average knowledge gain. However, the individual knowledge gain of the 
participants varied significantly (avg. standard deviation is $0.36$), 
reflecting the subjectivity of the learning process. 

Table~\ref{tab:kg-pearson-cor} shows the correlation between features and knowledge gain and the corresponding confidence level $\alpha$.
\begin{table}[h]
    \centering
	\begin{threeparttable}
		\begin{tabular}{|c|c|c|}
			\hline
			\textbf{Automatic Features}                    & \bm{$ r_{p} $}   & \boldmath{$\alpha<$} \\ \hline
			Modulated Loudness                   & -0.357          & 0.2   \\ \hline
			Image Ratio Var.                     & -0.282          & 0.3   \\ \hline 
			Coverage Avg.                        & 0.278           & 0.3   \\ \hline
		    Highlight                            & 0.264           & 0.3   \\ \hline
		    Coverage Var.                        & 0.253           & 0.3   \\ \hline
		    Speech Rate                          & -0.224          & 0.4   \\ \hline
		    Text Ratio Avg.                      & 0.220           & 0.4   \\ \hline
		    Avg. Syllable Duration               & 0.218           & 0.4   \\ \hline
		    Detailing Var.                       & -0.211          & 0.4   \\ \hline
		    RMS Energy                           & 0.209           & 0.4   \\ \hline
		    Harmonicity                          & 0.193           & 0.4   \\ \hline
		\end{tabular}
	\caption {Correlations between automatic features and results of the Knowledge Gain tests sorted by magnitude $ r_{p} $ with $\alpha<0.4$: Pearson Correlation Coefficient)}
	\label{tab:kg-pearson-cor}
	\end{threeparttable}  
\end{table}
The feature with the biggest positive impact on the knowledge gain with a value of ($r_{p}=-0.357,\alpha<0.2$) is \textit{Modulated Loudness}. This reflects the intuitive assumption that loud voices or unpleasant background noises hinder learning experience and effect. 
Knowledge gain also correlates negatively with \textit{Image Ratio Variance} ($r_{p}=-0.282,\alpha<0.3$), i.e., when the number of images varies significantly from slide to slide. This result indicates that a homogeneous slide design is to be preferred in a good presentation.
Another intuitive result is that \textit{Highlight of important Statements} positively correlates with knowledge gain ($r_{p}=0.264,\alpha<0.3$). Presumably, the process of highlighting the important statements helps the learner to focus on what to memorize and recollect the information later. Also, our second multimodal feature \textit{Coverage of Slide Content} ($r_{p}=0.278,\alpha<0.3$) showed a positive effect on the knowledge gain, confirming our intended design of these two metrics.
Another (weaker) finding is that \textit{speech rate} ($r_{p}=-0.224,\alpha<0.4$) and \textit{average syllable duration} ($r_{p}=0.218,\alpha<0.4$) are negatively correlated with the knowledge gain of the learner, representing the situation when a speaker talks too fast. The remaining features, even though they correlate only slightly, showed effects as expected. For instance, \textit{RMS Energy} and \textit{Harmonicity}, which measure the liveliness of the speaker's voice, had a positive correlation with the knowledge gain. 
\section{Conclusions}
\label{sec:conclusion}

In this paper, we have presented a set of unimodal and cross-modal features that can be automatically extracted from lecture videos. Further, we have presented a user study that investigated the correlations between these features and quality aspects of the lecture video. Also, the knowledge gain of users was measured and the correlations to the video features were evaluated. The results provided insights about a number of moderate to good correlations. 

We were able to represent the objective quality metrics \textit{Clear Language}, \textit{Vocal Diversity}, \textit{Speed Presentation} and \textit{Filler Words} each with at least three automatically extracted features and a confidence level of at least 95\%.
Additionally, we presented an approach to evaluate the level of \textit{Appropriate Detailing} which correlated with human assessment. 

In the future, we will investigate if and how these results generalize to other areas. 

Also, it is possible to measure the impact of user-related features like click behavior or gaze tracking data on the learning process and their correlation to our quality metrics.

Finally, we plan to utilize machine learning methods in order to predict video quality based on our set of features. This information could serve as an additional input for a recommender or retrieval system and add another dimension to distinguish between videos of similar content but differing presentation quality. In this way, exploratory search capabilities of educational video portals and MOOCs could be improved.

\begin{acks}
Part of this work is financially supported by the Leibniz Association, Germany (Leibniz Competition 2018, funding line "Collaborative Excellence", project SALIENT [K68/2017]).
\end{acks}

\newpage
\bibliographystyle{ACM-Reference-Format}
\bibliography{references}

\end{document}